\documentclass[reprint,prb]{revtex4-2}
\usepackage{amsmath,amssymb,amsfonts}
\usepackage{graphicx}
\usepackage[colorlinks=true,citecolor=blue]{hyperref}

\begin{document}

\title{Global optimization of tensor renormalization group using the corner transfer matrix}
\author{Satoshi Morita}
\email[]{morita@issp.u-tokyo.ac.jp}
\author{Naoki Kawashima}
\affiliation{Institute for Solid State Physics, University of Tokyo,
Kashiwa, Chiba 277-8581, Japan}
\date{\today}

\begin{abstract}
A tensor network renormalization algorithm with global optimization based on the corner transfer matrix is proposed.
Since the environment is updated by the corner transfer matrix renormalization group method, the forward-backward iteration is unnecessary, which is a time-consuming part of other methods with global optimization.
In addition, a further approximation reducing the order of the computational cost of contraction for the calculation of the coarse-grained tensor is proposed.
The computational time of our algorithm in two dimensions scales as the sixth power of the bond dimension while the higher-order tensor renormalization group and the higher-order second renormalization group methods have the seventh power.
We perform benchmark calculations in the Ising model on the square lattice and show that the time-to-solution of the proposed algorithm is faster than that of other methods.
\end{abstract}

\maketitle

\section{Introduction}

Tensor network methods attract much attention as powerful tools for computing strongly correlated many-body problems \cite{Orus_review,Ran2020}.
The partition function of classical statistical systems and low-energy states in quantum systems can be represented by tensor networks, in which exponentially large information is compressed efficiently.
However, contraction of a large tensor network still requires a huge computational cost.
The combination of tensor networks and real-space renormalization group ideas resolves this problem.
The tensor renormalization group (TRG) method provide a way to calculate a coarse-grained tensor based on the singular value decomposition \cite{TRG}.
The truncation of smaller singular values avoids divergence of tensor size.
The higher-order tensor renormalization group (HOTRG) method is another method applicable to higher-dimensional systems \cite{HOTRG}.

Both of the methods do information compression by solving local optimization problems.
Approximations in these methods are locally optimal, but not so for contraction of the whole tensor network.
Therefore, the global optimization is necessary.
Since the global optimization problem is defined by using the whole network, we need to introduce other approximations.
The whole network is split into two parts, i.e., a small system and an environment surrounding it, and then the latter is approximated in an appropriate way.
The second renormalization group (SRG) method \cite{SRG2009,SRGfull2010} and the higher-order SRG (HOSRG) method \cite{HOTRG} represent an environment as one tensor, which is called the environment tensor.
Recently, the automatic differentiation technique was proposed to calculate the environment tensor \cite{Chen2019}.
Although these methods drastically improve accuracy, calculation of the environment tensor requires performing the forward-backward iterations.

In this paper, we propose another approximation of the environment.
We replace the environment tensor with the corner transfer matrices (CTMs) \cite{Baxter_book} and the edge tensors, which we call the CTM environment.
It can be updated by using the CTM renormalization group (CTMRG) \cite{CTMRG1996,CTMRG1997} instead of the backward iteration.
The computational cost of the CTMRG is smaller than the backward iteration.
The former scales as $O(\chi^6)$ against the bond dimension $\chi$ while the latter has $O(\chi^7)$ scaling.
In addition, we introduce an additional decomposition, whose key idea is information compression with the environment.
This approximation reduces the computational cost of tensor contraction for the coarse-grained tensor.
Finally, our algorithm achieves computational cost scales as $O(\chi^6)$, while HOTRG and HOSRG have $O(\chi^7)$ cost in two-dimensional systems.

In the next section, we introduce our improved algorithm, which we call CTM-TRG.
In the third section, benchmark results performed in the two-dimensional Ising model are shown.
We will show that our algorithm has a smaller time-to-solution than HOTRG and HOSRG.
The last section is devoted to discussions and conclusions.

\section{Algorithm}

Let us consider the contraction of a tensor network on the square lattice,
\begin{equation}
  Z = \text{Cont}\left( \prod_{i} T_i \right).
\end{equation}
A local tensor $T_i$ located at each site $i$ has four indices and connects with other tensors on the nearest-neighbor sites.
We assume that each index of $T$ runs from $1$ to $\chi$ at most.
In other words, the bond dimension of $T$ is equal to $\chi$.
For classical systems, $Z$ is the partition function and the Boltzmann weight determines elements of $T_i$.
In quantum systems, such a contraction commonly appears as an inner product of a wave function, the so-called tensor network state or the projected entanglement paired state~\cite{PEPS2004}.

\begin{figure}
  \includegraphics[width=\columnwidth]{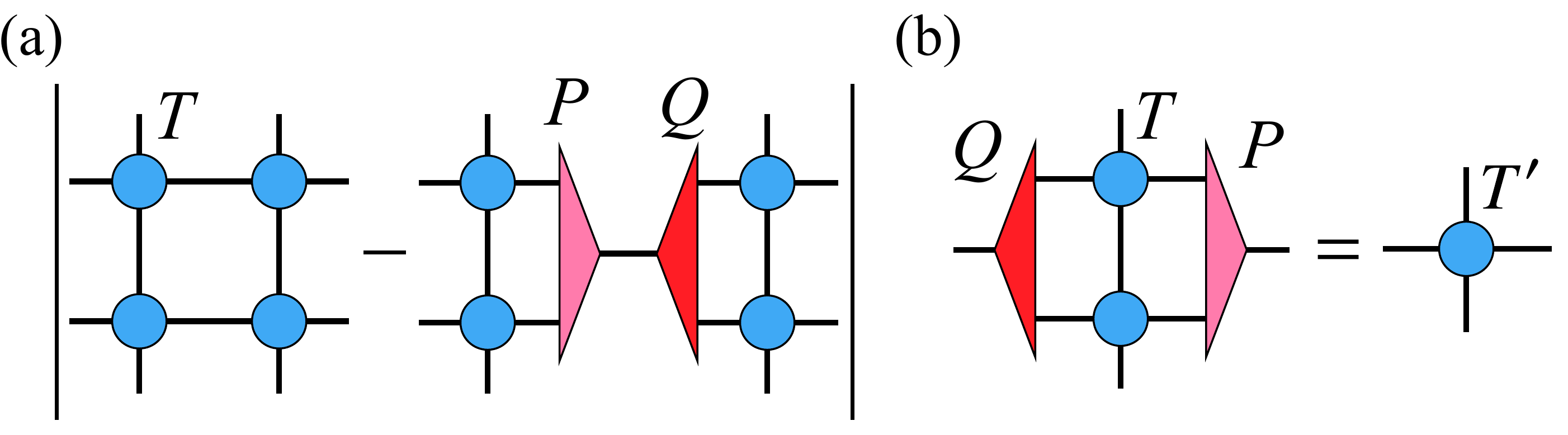}
  \caption{\label{fig:HOTRG} (Color online)
  Graphical representations of the HOTRG algorithm.
  (a) The bond-merging operators in HOTRG are given as a solution of the local optimization problem to minimize this norm.
  (b) A coarse-grained tensor $T'$ is obtained by contraction.}
\end{figure}

The HOTRG algorithm approximates this contraction by introducing the operator which merges two bonds into a single bond.
In the original paper of HOTRG~\cite{HOTRG}, this bond-merging operator was obtained by the higher-order singular value decomposition (HOSVD) of a tensor,
\begin{equation}
  M_{(x_1x_2)y(x'_1x'_2)y'} \equiv \sum_{i}
  T_{x_1yx'_1i} T_{x_2ix'_2y'}.
\end{equation}
Another solution of the bond-merging operator is the oblique projector $PQ$ which minimizes
\begin{equation}
  \left\| M M - M P Q M \right\|
  \label{eq:local_probrem}
\end{equation}
while keeping the bond dimension between $P$ and $Q$ to $\chi$.
Its graphical representation is shown in Fig.~\ref{fig:HOTRG}(a).
The algorithm for the calculation of $PQ$ is well known~\cite{Wang2011, CorbozTroyer2014, loopTNR, BTRG} and its computational cost scales as $O(\chi^6)$~\cite{BTRG}.
After inserting bond-merging operators, the coarse-grained tensor is defined by contraction,
\begin{equation}
  T'_{xyx'y'} = \sum_{x_1x_2x'_1x'_2}
  M_{(x_1x_2)y(x'_1x'_2)y'} P_{(x_1x_2)x} Q_{x'(x'_1x'_2)}
\end{equation}
The computationally heaviest part in HOTRG is this tensor contraction [Fig.~\ref{fig:HOTRG}(b)], which scales as $O(\chi^7)$.

\begin{figure}
  \includegraphics[width=\columnwidth]{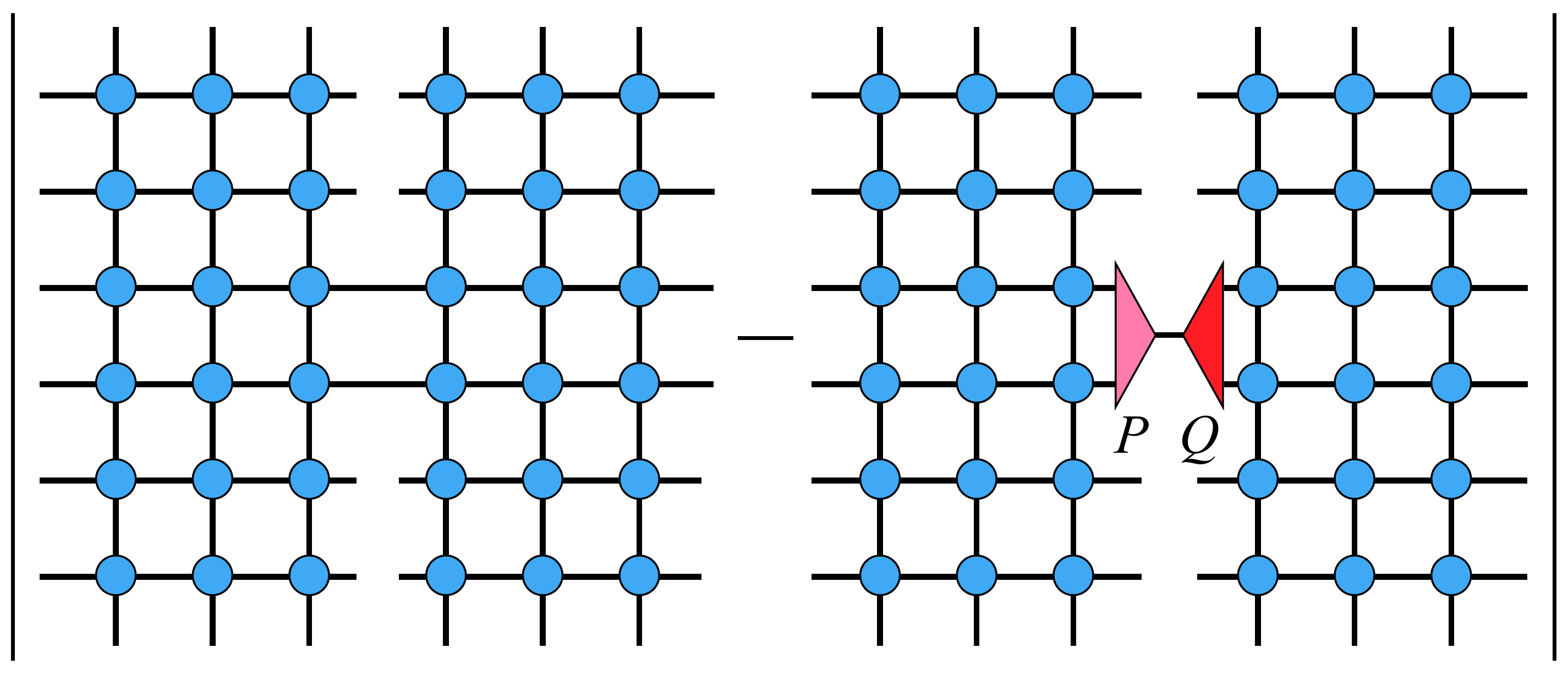}
  \caption{\label{fig:cluster_opt} (Color online)
  More accurate bond-merging operators $P$ and $Q$ are obtained as a solution of the optimization problem with larger tensor networks.
  Here, the cost function of the optimization problem with $6\times 6$ clusters is shown.}
\end{figure}

The minimization problem for the bond-merging operator in HOTRG (\ref{eq:local_probrem}) is a local optimization problem.
Although such a solution is the best for a $2\times 2$ cluster, it is not the case for contraction of the whole network.
A bond-merging operator obtained from larger clusters such as in Fig.~\ref{fig:cluster_opt} will improve the accuracy and its limit to infinite cluster size converges to a solution of the global optimization problem which minimizes the difference between the whole network with and without a bond-merging operator.
However, an optimization problem with large clusters is intractable because its computational cost rapidly diverges with the cluster size.

Since the global optimization problem involves contraction of the whole network, some approximations are necessary.
In the HOSRG algorithm, one local tensor is picked up and the other part is considered as an environment.
The environment is represented by a four-index tensor, which we call the environment tensor.
More precisely, the local tensor $T^{(t)}$ and corresponding environment tensor $\text{Env}^{(t)}$ at each renormalization step $t$ approximate the contraction of the whole network as
\begin{equation}
  Z \simeq \sum_{x,y,x',y'} T^{(t)}_{xyx'y'} \text{Env}^{(t)}_{xyx'y'},
  \label{eq:env}
\end{equation}
The calculation of the environment tensor is done by the so-called backward iteration.
It is a fine-grained process which updates the environment tensor $\text{Env}^{(t-1)}$ using information at the $t$-th step.
On the other hand, a coarse-grained process, called the forward iteration, updates the local tensors and the bond-merging operators.
The HOSRG algorithm repeats the forward and backward iterations until convergence.

We note that the computational cost of the HOSRG algorithm scales as $O(\chi^7)$, as does HOTRG.
Although calculation of the bond density operator defined in Ref.~\cite{HOTRG} requires $O(\chi^8)$ computational cost, it can be avoided by using a similar technique shown in Ref.~\cite{TRG_RSVD}.
Thus, both the forward and backward iterations have $O(\chi^7)$ computational cost.

The environment tensor requires the backward iteration and it causesthe  main difficulty of the HOSRG method.
The CTM-TRG employs the corner transfer matrix and the edge tensor since these can be calculated only from the local tensor.
The environment tensor is decomposed into four CTMs $\{C_\text{TR}, C_\text{BR}, C_\text{BL}, C_\text{TL}\}$ and four edge tensors $\{E_\text{T}, E_\text{L}, E_\text{B}, E_\text{R}\}$, which we call the CTM environment in this paper.
The subscripts (T, R, B, and L) indicate the positions from a local tensor (top, right, bottom, and left, respectively).
For example, the top-right CTM $C_\text{TR}$ represents all the tensors in the first quadrant.

\begin{figure}
  \includegraphics[width=\columnwidth]{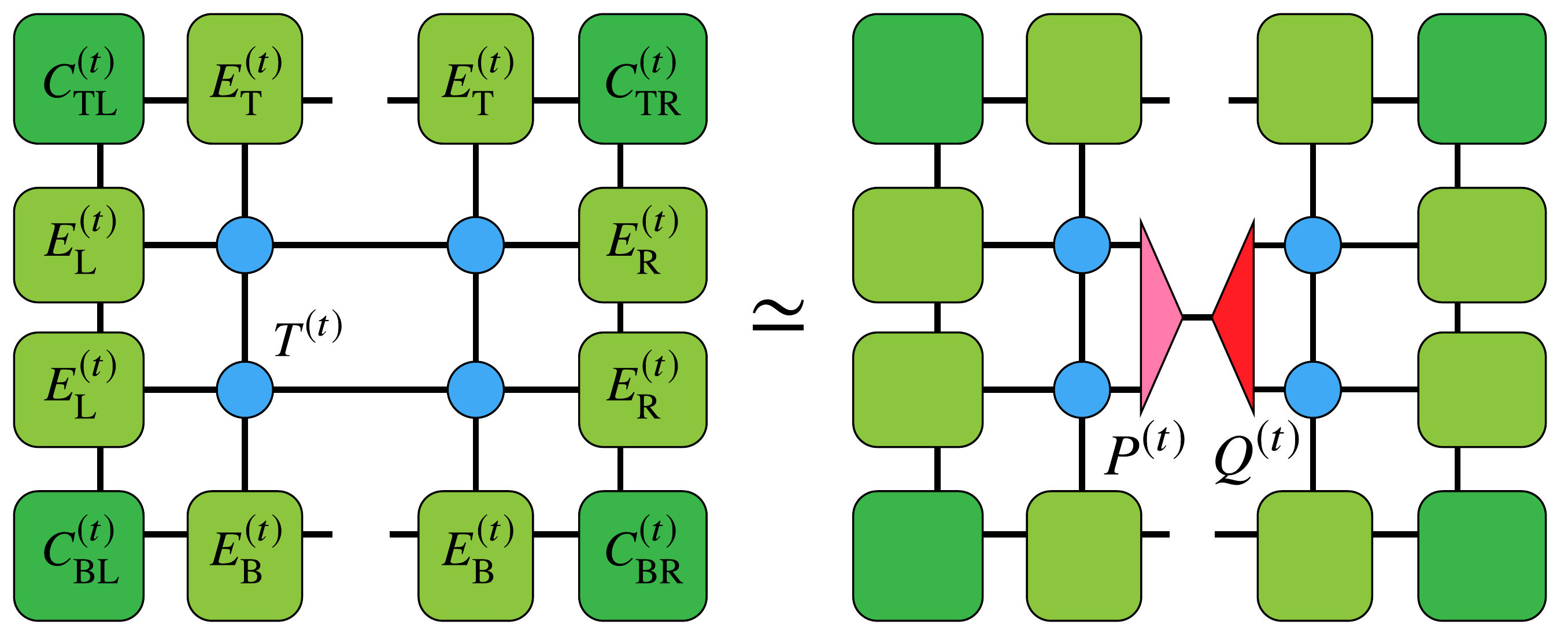}
  \caption{\label{fig:CTMTRG_proj} (Color online)
  The bond-merging operators in CTM-TRG are obtained as the oblique projector between the left- and right-half networks.}
\end{figure}

The global optimization problem is represented by a $2\times2$ cluster of local tensors and its surrounding CTM environment.
The bond-merging operator is calculated as the oblique projector between two half networks as shown in Fig.~\ref{fig:CTMTRG_proj}.
Both the contraction of the half network and calculation of the bond-merging operator have a computational cost proportional to $\chi^6$.

\begin{figure}
  \includegraphics[width=2.5in]{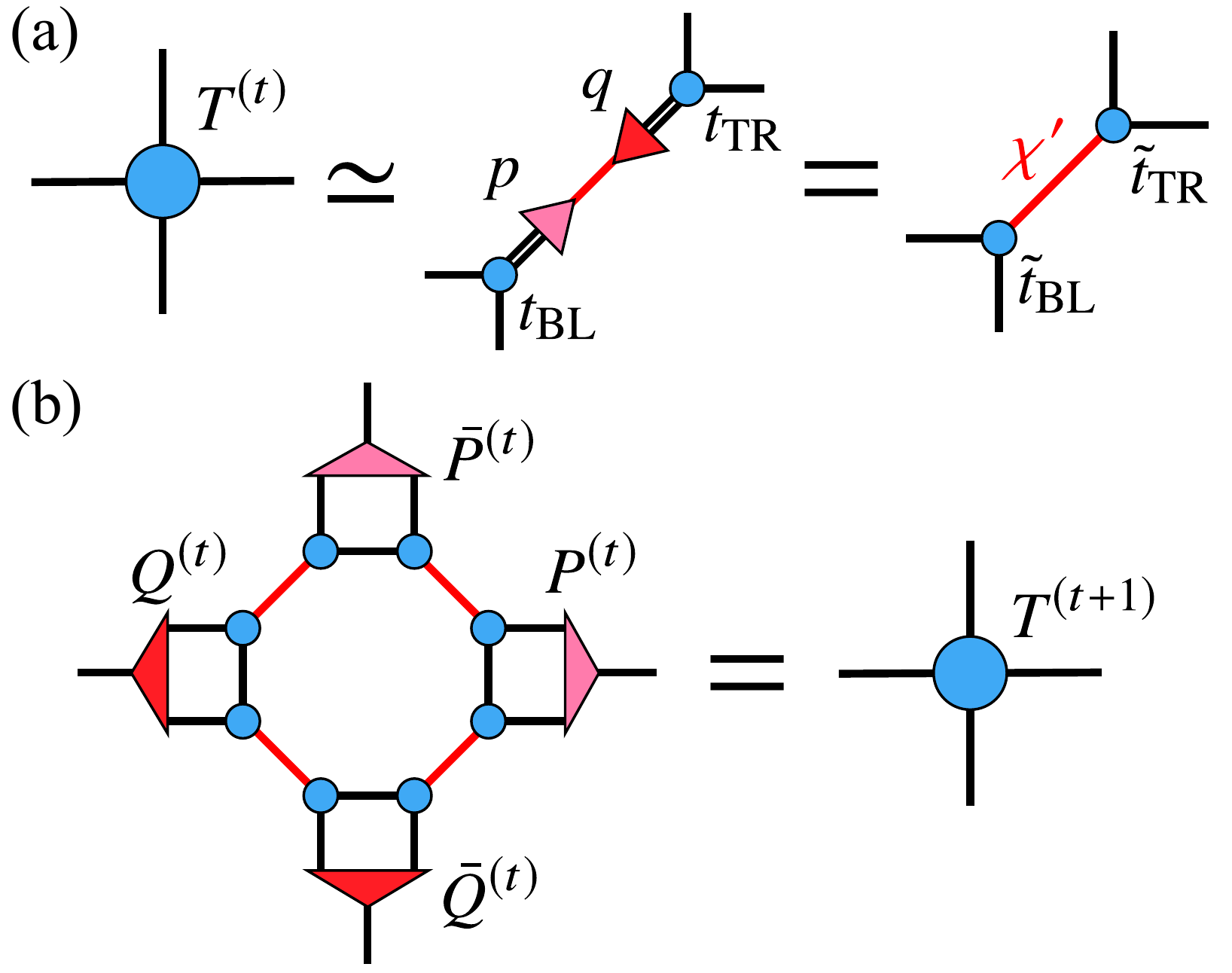}
  \caption{\label{fig:CTMTRG_cont2} (Color online)
  Graphical representation of the additional approximation, which reduces the computational cost of contraction for the coarse-graining tensor. The red diagonal lines indicate that their bond dimension is $\chi'$.
  (a) The local tensor is decomposed into two three-index tensors.
  The truncation operators $p$ and $q$ reduce the bond dimension from $\chi^2$ to $\chi'$ (see also Fig.~\ref{fig:CTMTRG_cont3}).
  (b) A coarse-grained tensor in CTM-TRG is obtained by the simultaneous scale transformation along the horizontal and vertical directions.}
\end{figure}

\begin{figure}
  \includegraphics[width=2.5in]{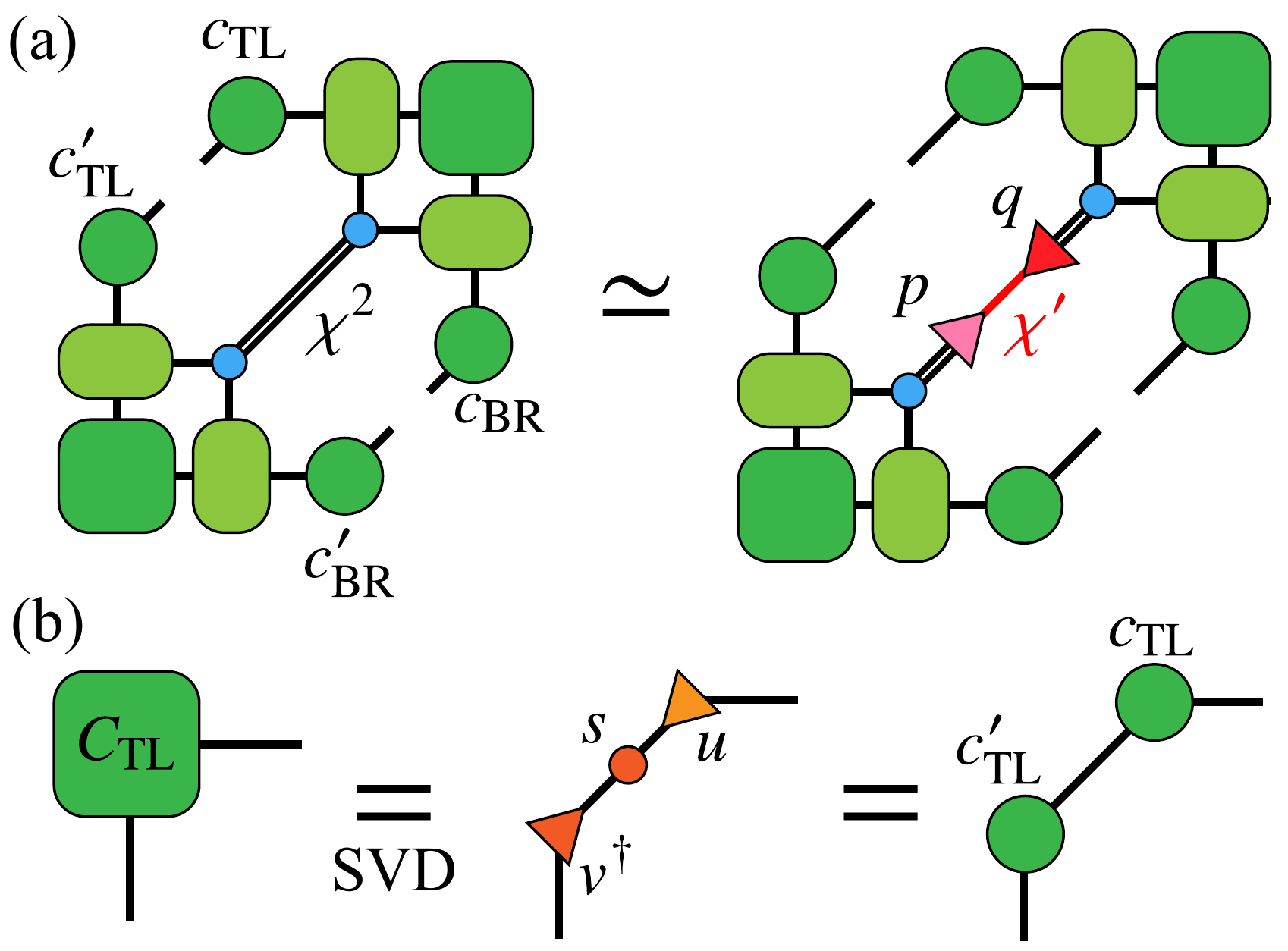}
  \caption{\label{fig:CTMTRG_cont3} (Color online)
  (a) The truncation operators $p$ and $q$ in Fig.~\ref{fig:CTMTRG_cont2}(a) are calculated as the oblique projector which minimizes the difference between the two networks.
  (b) The corner matrix $C_\text{TL}$ is decomposed into $c_\text{TL}$ and $c'_\text{TL}$ by using the singular value decomposition.
  }
\end{figure}

The CTM environment reduces the computational cost for the calculation of the bond-merging operator to $O(\chi^6)$.
However, contraction for the coarse-grained tensor [Fig.~\ref{fig:HOTRG}(b)] still scales as $O(\chi^7)$.
To reduce this cost, we introduce an additional approximation (Fig.~\ref{fig:CTMTRG_cont2}).
First, we decompose the local tensor by using the singular-value decomposition,
$T_{xyx'y'}=\sum_{a=1}^{\chi^2} U_{xya}\Sigma_a V^{*}_{x'y'a}$,
and define $t_\text{TR}=U\sqrt{\Sigma}$ and $t_\text{BL}=\sqrt{\Sigma}V^{\dagger}$.
Next, we insert truncation operators $p$ and $q$ between $t_\text{TR}$ and $t_\text{BL}$ to reduce the bond dimension from $\chi^2$ to $\chi'$.
These operators are also calculated as oblique projectors which minimize the difference shown in Fig.~\ref{fig:CTMTRG_cont3}(a), where $c_\text{TL}$ and $c'_\text{TL}$ ($c_\text{BR}$ and $c'_\text{BR}$) are obtained from the singular value decomposition of $C_\text{TL}$ ($C_\text{BR}$) such as
$C_\text{TL}= usv^{\dagger}$, $c_\text{TL}=u\sqrt{s}$, and $c'_\text{TL}=\sqrt{s}v^\dagger$ [Fig.~\ref{fig:CTMTRG_cont3}(b)].
We define $\tilde{t}_\text{TR}=t_\text{TR} p$ and $\tilde{t}_\text{BL}=q t_\text{BL}$.
In the same manner, we calculate $\tilde{t}_\text{TL}$ and $\tilde{t}_\text{BR}$.
Finally, we obtain the coarse-grained tensor $T^{(t+1)}$ by contraction of the tensor network as shown in Fig.~\ref{fig:CTMTRG_cont2}(b), where $\bar{P}$ and $\bar{Q}$ are the bond-merging operators for vertical bonds.
Clearly, the computational cost of each step scales as $O(\chi^6)$ at most.
In contrast to HOTRG and HOSRG, this method performs scale transformations along the horizontal and vertical axes simultaneously.

At the beginning of the next step, we need to update the CTM environment by using CTMRG with the local tensor $T^{(t+1)}$.
An initial value of the CTM environment at the step $t+1$ can be easily generated from the CTM environment at the previous step $t$.
An initial value of $C^{(t+1)}$ is equal to $C^{(t)}$ and that of $E^{(t+1)}$ is calculated from $E^{(t)}$ and the bond-merging operators.
For example, the edge tensor on the left edge is initialized as
\begin{equation}
  E^{(t+1)}_{\text{L}, xyy'} =
  \sum_{x_1,x_2,y_1}
  E^{(t)}_{\text{L}, x_1yy_1}
  E^{(t)}_{\text{L}, x_2y_1y'}
  P^{(t)}_{x_1x_2x}.
\end{equation}
Since these tensors are a good initial guess for the CTM environment at $t+1$, the number of CTMRG iterations for an update of the CTM environment is smaller than that for initialization of the CTM environment at $t=0$.
We note that this contraction has only $O(\chi^5)$ computational cost.

\section{Results}

We simulate the Ising model on the square lattice to investigate the performance of our proposed algorithm.
The initial local tensor at temperature $T=1/\beta$ is given as
\begin{equation}
  T_{xyx'y'}^{(0)} = \sum_{s=1,2} W_{sx}W_{sy}W_{sx'}W_{sy'},
\end{equation}
where $W$ is a $2\times 2$ matrix,
\begin{equation}
  W = \begin{pmatrix}
    \sqrt{\cosh \beta} & \sqrt{\sinh \beta}\\
    \sqrt{\cosh \beta} & -\sqrt{\sinh \beta}\\
  \end{pmatrix}.
\end{equation}
The critical temperature is $T_c = 2/\ln(\sqrt{2}+1)$.
The free energy per site is estimated from the coarse-grained tensor $T^{(t)}$ as
\begin{equation}
  f = -\frac{1}{\beta N} \ln \sum_{xy} T_{xyxy}^{(t)},
  \label{eq:def_f}
\end{equation}
where $t$ is the number of renormalization steps and $N=2^{2t}$.
We perform 20 renormalization steps, which is enough to observe convergence of the free energy even when its relative error is less than $10^{-12}$.
Because of the global optimization, the free energy in CTM-TRG converges to its value in the thermodynamic limit much faster than that in HOTRG except in the near-critical region.
We note that the free energy can also be estimated from the CTMs~\cite{CTMRG1996}.
It corresponds to the fixed or open boundary condition, while Eq.~(\ref{eq:def_f}) assumes the periodic boundary condition.
The both should take the same value in the thermodynamic limit and we confirm this fact in our numerical simulations.

In our algorithm, we use the same bond dimension $\chi$ for the local tensor and CTM environment and set $\chi'=2\chi$ for the diagonal decomposition of a local tensor in Fig.~\ref{fig:CTMTRG_cont2}.
In most cases, we use the fixed boundary condition (FBC) for the initial condition of the CTM environment.
The initial edge tensor $E^{(0)}$ is a $2\times 1 \times 1$ tensor and its element is given as $E_{x11} = W_{1x}$, where the first index connects with a local tensor.
The initial CTM is a $1\times 1$ identity matrix, that is, $C_{11}=1$.
We first perform $32$ CTMRG steps to obtain the CTM environment at $t=0$ and do four CTMRG steps per each update of $T^{(t)}$.
Although we observe that four is not enough to achieve convergence near the criticality, it is still sufficient for making the whole procedure produce more accurate results than HOTRG.
We also use the open boundary condition (OBC), which is expected to be better than FBC in the paramagnetic phase.
The initial CTM and edge tensor for OBC are defined by using $W$ as well as the initial local tensor $T^{(0)}$, for example, $C_{xy}=\sum_{s}W_{sx}W_{sy}$.
For comparison, we also perform HOTRG and HOSRG simulations.
In HOSRG, we repeat the forward-backward iterations four times because of our observation that it is sufficient for the convergence of the free energy in all the cases.

\begin{figure}
  \includegraphics[width=\columnwidth]{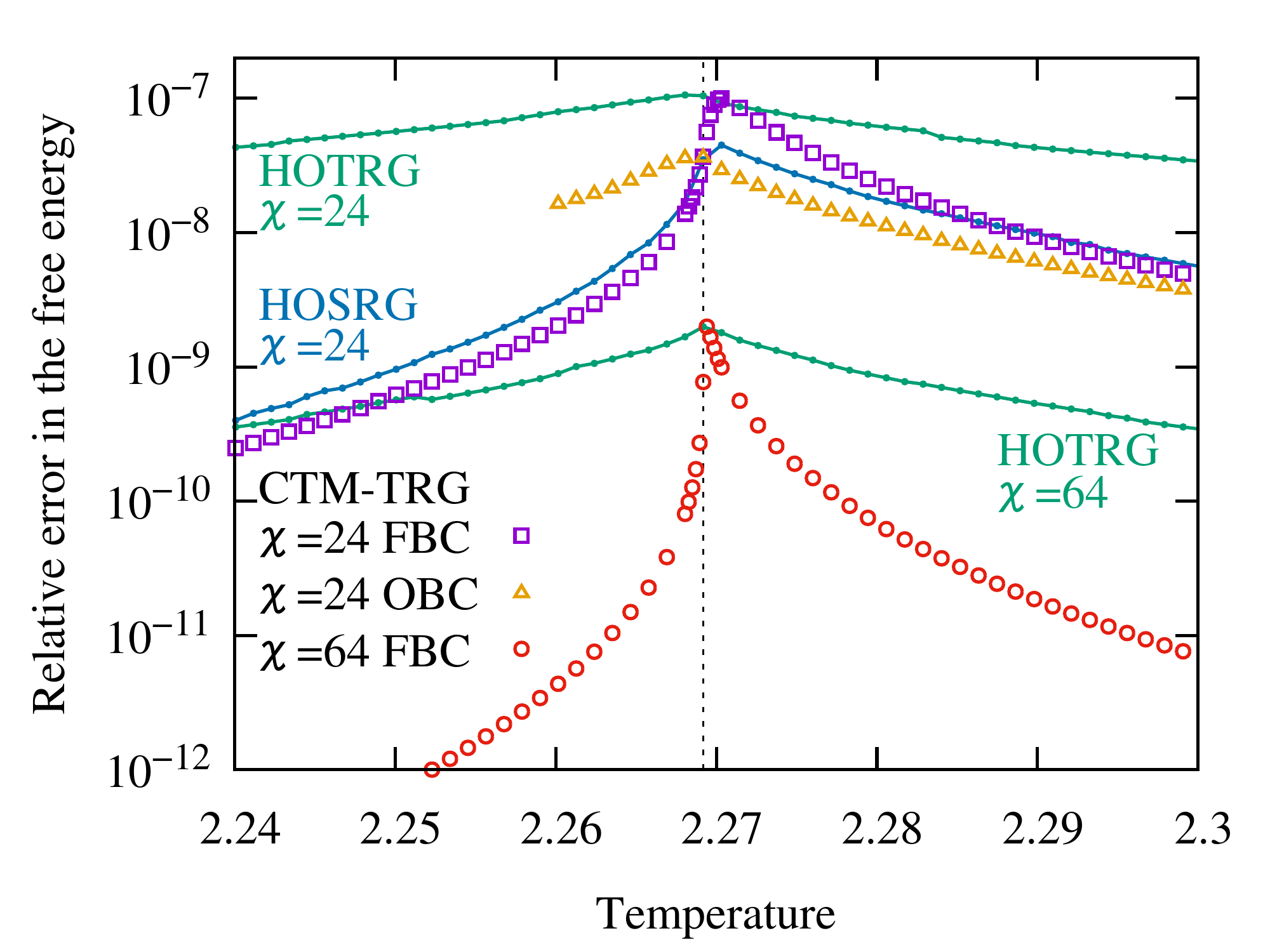}
  \caption{\label{fig:f_vs_temp} (Color online)
  Comparison of relative errors in the free energy.
  The vertical dashed line indicates the critical temperature.
  }
\end{figure}

We compare the relative errors in the free energy from the exact solution in the thermodynamic limit (Fig.~\ref{fig:f_vs_temp}).
The method of CTM-TRG shows better accuracy than HOTRG in all cases, and is compatible to HOSRG.
In the ferromagnetic phase, CTM-TRG is more accurate than HOSRG.
This is because CTM-TRG performs scale transformations along horizontal and vertical directions simultaneously [Fig.~\ref{fig:CTMTRG_cont2}(b)].
We confirmed that our algorithm without the $O(\chi^6)$ approximation shows the same accuracy as HOSRG.
Above the critical temperature, CTM-TRG has a slightly larger error than HOSRG, which is caused by the small number of CTMRG iterations.
The open boundary condition is more suitable for an initial value of the CTM environment in the paramagnetic phase and it provides more accurate results than HOSRG.
We note that the results with $\chi=64$ are much better than the results reported in Ref.~\cite{Chen2019}, in which a one-dimensional renormalization algorithm for a large bond dimension was performed.

\begin{figure}
  \includegraphics[width=\columnwidth]{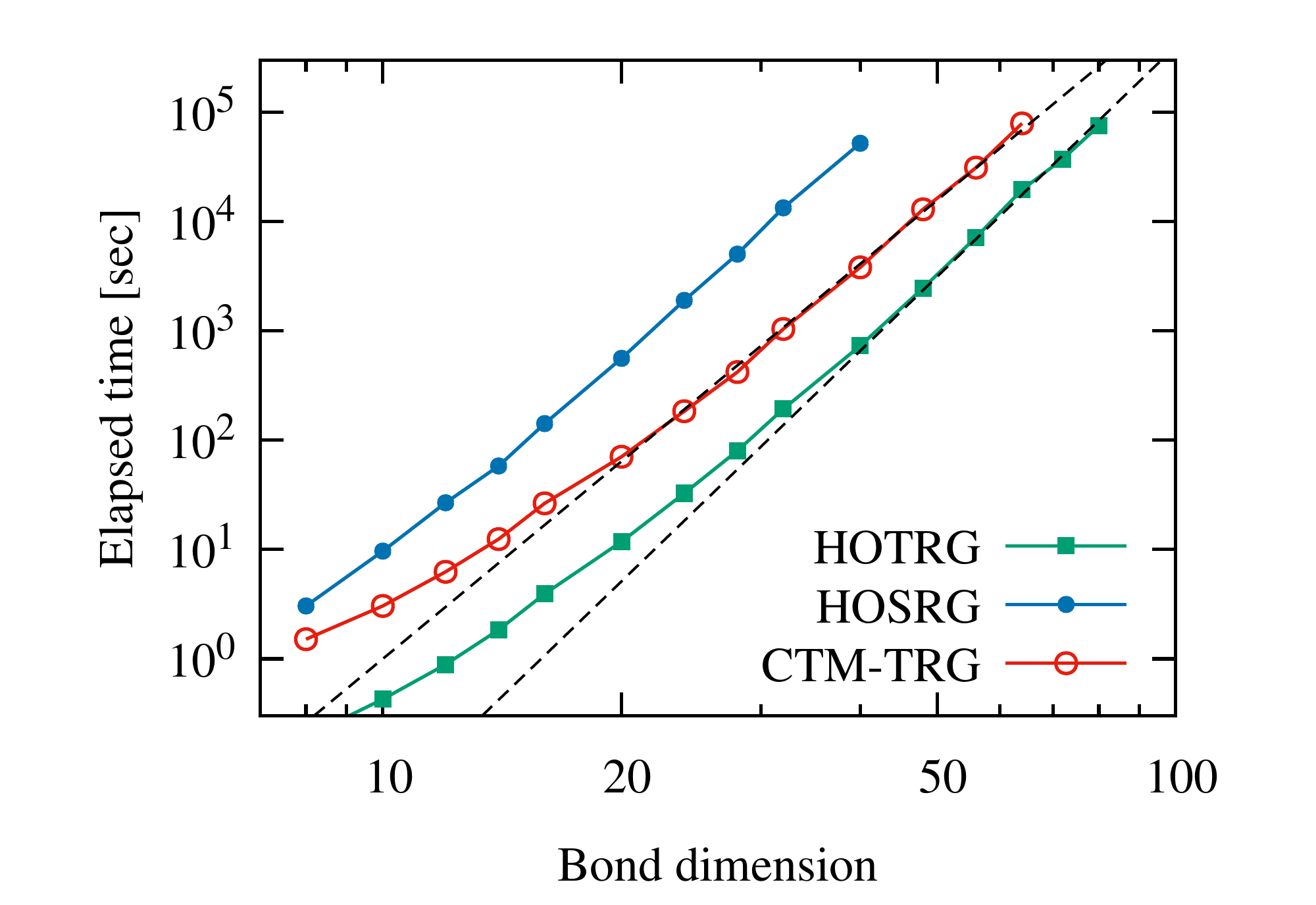}
  \caption{\label{fig:time_vs_chi} (Color online)
  Elapsed time of 20 renormalization steps as a function of the bond dimension. The dashed lines proportional to $\chi^6$ and $\chi^7$ are guides for the eye.
  }
\end{figure}

The elapsed time of each method is shown in Fig.~\ref{fig:time_vs_chi}.
Our data clearly show $O(\chi^6)$ scaling of CTM-TRG while HOTRG and HOSRG have $O(\chi^7)$.
Although the elapsed time of CTM-TRG is longer than HOTRG in the range of the bond dimension we calculated, they will switch places around $\chi=250$.
The computational time was measured by simulations in a single core on Intel Xeon E5-2697A (2.60 GHz) with 128 GB memory.

\begin{figure}
  \includegraphics[width=\columnwidth]{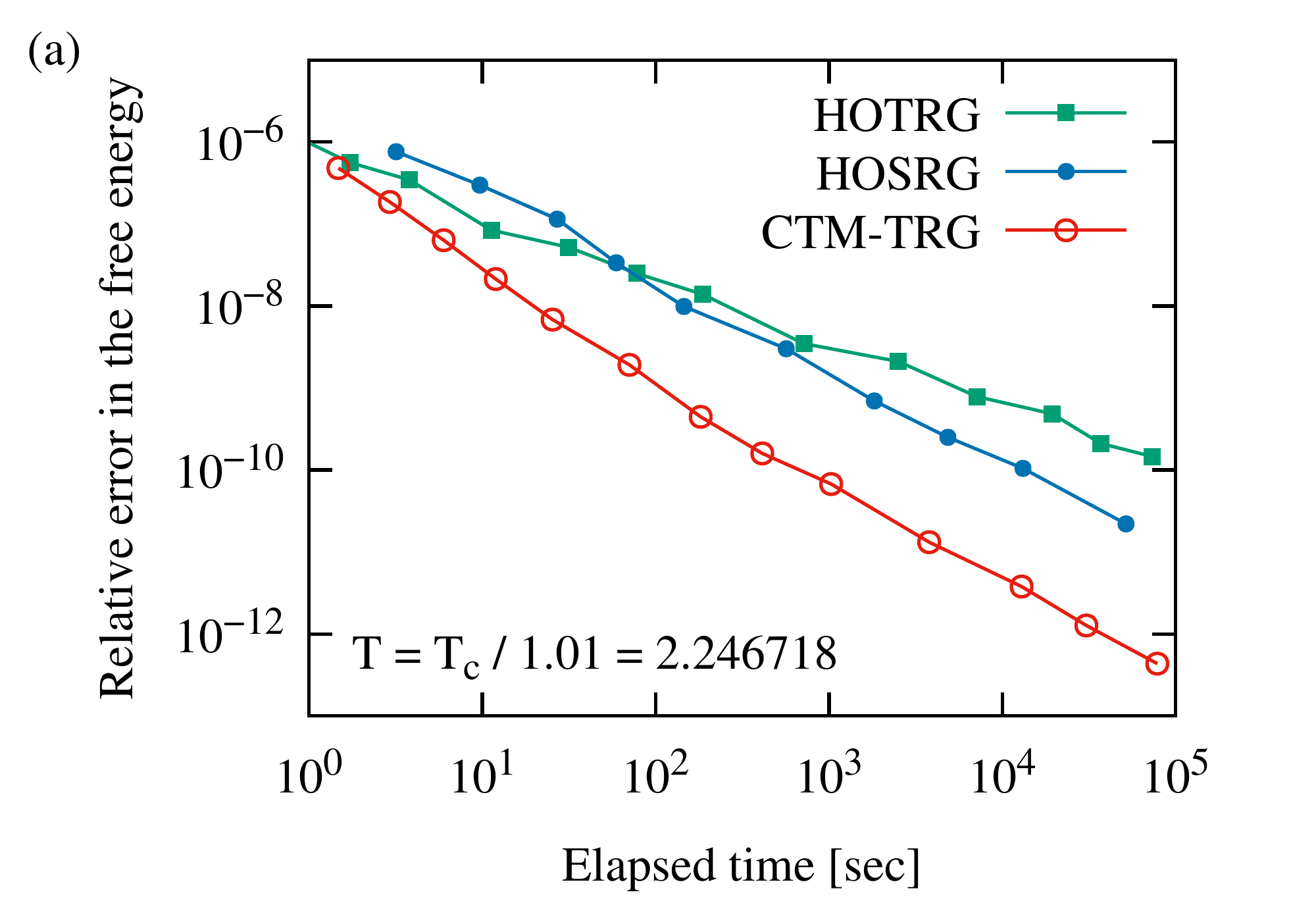}
  \includegraphics[width=\columnwidth]{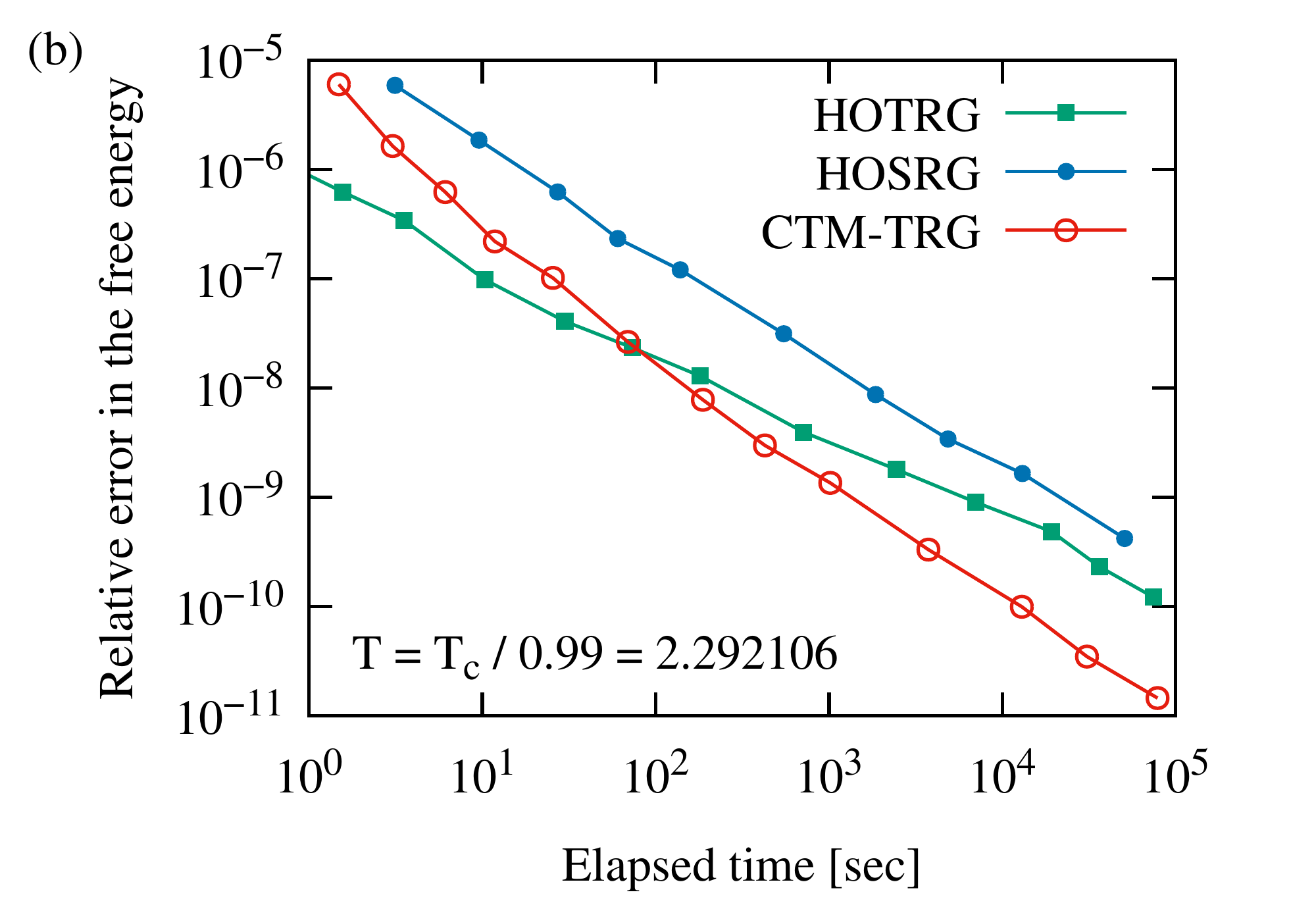}
  \caption{\label{fig:f_vs_time} (Color online)
  Relative error in the free energy at (a) $T=T_c/1.01$ and (b) $T_c/0.99$ as a function of the elapsed time of 20 steps.
  }
\end{figure}

Figure \ref{fig:f_vs_time} shows the time-to-solution at $T=T_c/1.01$ and $T_c/0.99$.
CTM-TRG always outperform the other methods at $T=T_c/1.01$.
This is because the global optimization drastically improves accuracy in the ferromagnetic phase as shown in Fig.~\ref{fig:f_vs_temp}.
In the paramagnetic phase ($T=T_c/0.99$), CTM-TRG has the steepest slope and achieves the best performance with large bond dimensions.

\begin{figure}
  \includegraphics[width=\columnwidth]{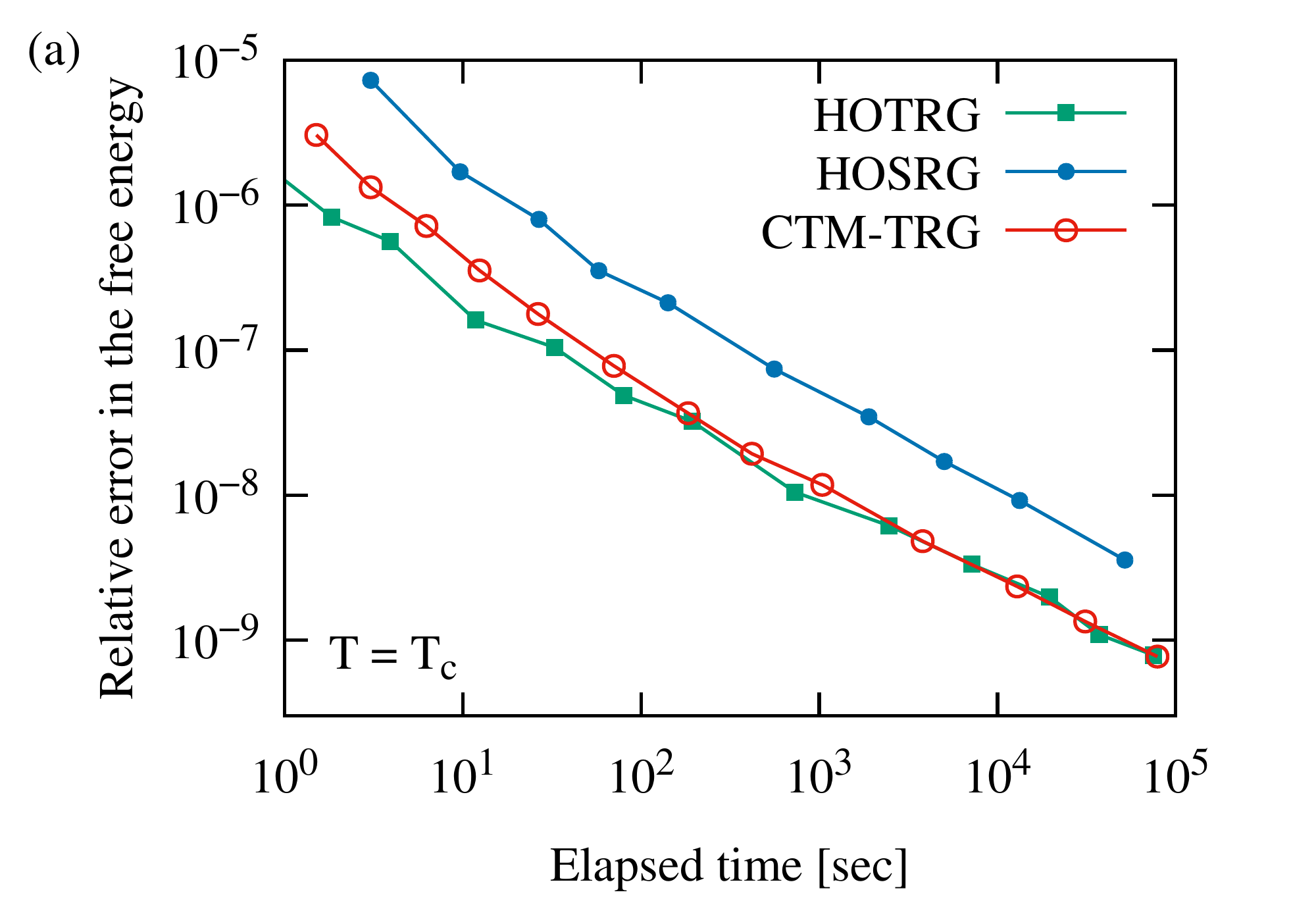}
  \includegraphics[width=\columnwidth]{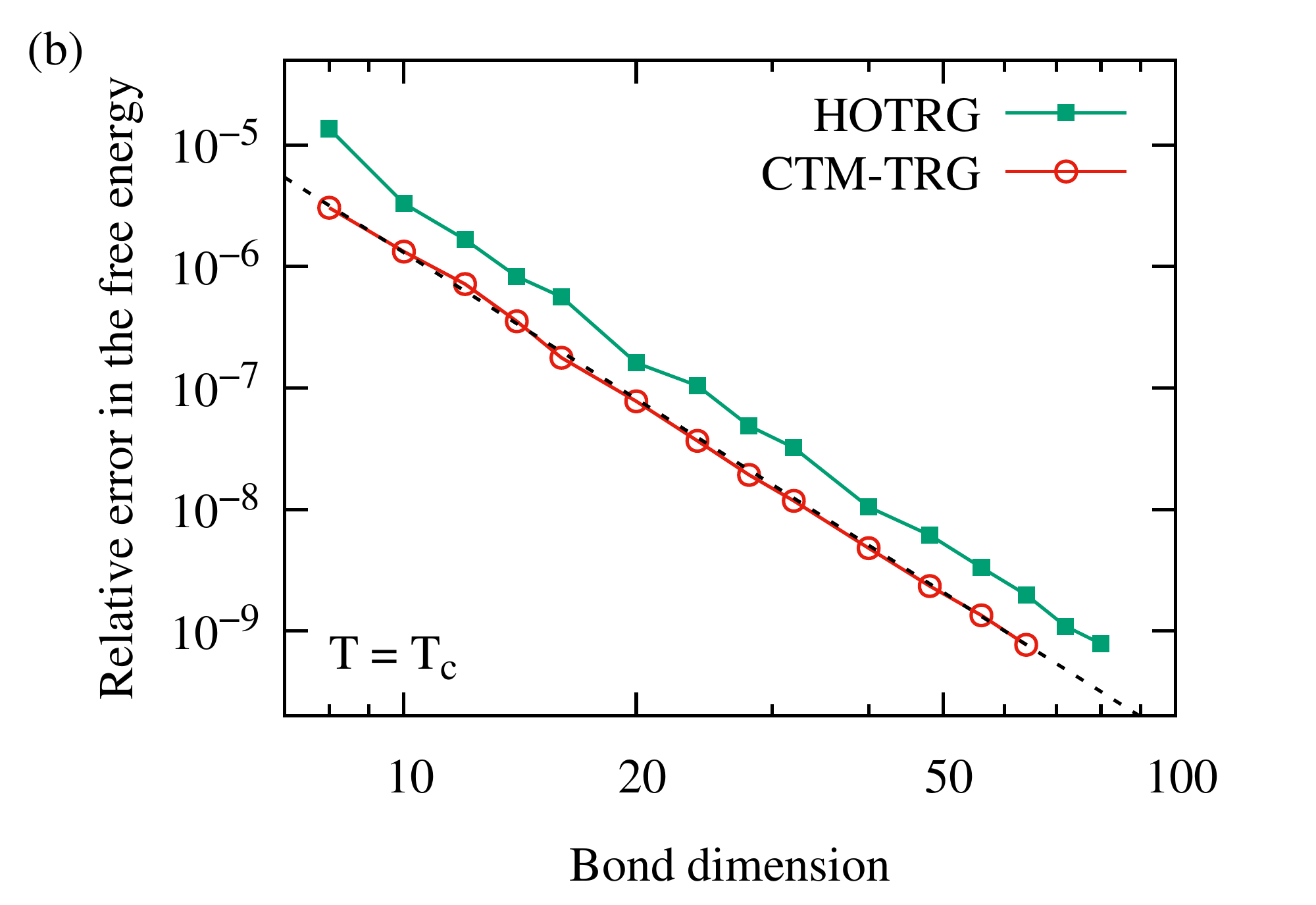}
  \caption{\label{fig:f_vs_time_Tc} (Color online)
  Relative error in the free energy at criticality as a function of (a) the elapsed time and (b) the bond dimension.
  The dashed line shows a fitting result.
  }
\end{figure}

In Fig.~\ref{fig:f_vs_time_Tc}, we show the performance of CTM-TRG at criticality.
Although CTM-TRG and HOTRG seem to have almost the same time-to-solution, the former should be superior to the latter in larger bond dimensions.
As shown by the dashed line in Fig.~\ref{fig:f_vs_time_Tc}(b), we find that the relative error in the free energy of CTM-TRG is $\varepsilon_f \sim 0.013\times \chi^{-4.0}$ and CTM-TRG achieves the same accuracy with 20\% smaller bond dimension than HOTRG.
From these facts and scaling of the computational time, we estimate that the crossing point between CTM-TRG and HOTRG exists at about $20000$ seconds, as shown in Fig.~\ref{fig:f_vs_time_Tc}(a).
The relative error of CTM-TRG and HOTRG is proportional to  $\tau^{-0.66}$ and $\tau^{-0.57}$, respectively.

\section{Discussion and conclusions}

In this paper, we consider the global optimization of a tensor renormalization group and propose the improved algorithm based on the CTM environment.
In our proposed algorithm CTM-TRG, the environment tensor of the HOSRG method is replaced by the CTMs and edge tensors.
Since the CTM environment can be easily updated by using CTMRG, our algorithm does not require any backward iteration.
In addition, we introduce an additional approximation by decomposing the four-rank tensor into three-rank tensors which reduces the order of the computational cost for tensor contraction without a serious reduction in the overall accuracy.
The computational cost of each step in CTM-TRG is bounded by $O(\chi^6)$, while HOTRG and HOSRG have $O(\chi^7)$ computational cost.
Therefore, our algorithm can produce almost the same accuracy as HOSRG within a smaller computational time.
We also show that the time-to-solution of CTM-TRG is shorter than HOTRG and HOSRG in the paramagnetic and ferromagnetic phases.

In the present work, we use the standard CTMRG method to calculated the CTM environment~\cite{CTMRG1996,CorbozTroyer2014}.
Recently, a variational method based on the uniform matrix product state was proposed \cite{VUMPS} and it was applied to the contraction of two-dimensional tensor networks \cite{Fishman2018}.
The edge tensors construct an eigenvector of the row-to-row (or column-to-column) transfer matrix represented as a matrix product operator, and the CTM is calculated as a solution of the fixed point equation.
This algorithm improves the convergence speed of the CTM environment, especially near the critical point.
Thus usage of such a variational approach instead of CTMRG may improve the performance of our proposed algorithm.

We also comment on the applicability of CTM-TRG to higher-dimensional systems.
Although we have focused on the two-dimensional systems in this paper, the global optimization using the CTM environment can be generalized to higher dimensions.
In three dimensions, the face tensor in addition to the CTMs and edge tensors is necessary to construct the environment~\cite{CTMRG3D}.
The CTM calculation in high dimensions is more difficult than that in two dimensions because of the slow decay of the singular values.
We may need to use additional techniques to obtain accurate results as mentioned below.

It is known that TRG and HOTRG may converge to a fictitious fixed point owing to short-range entanglement~\cite{TEFR,Ueda2014},
and the calculation of the scaling dimensions from eigenvalues of the transfer matrix fails~\cite{TEFR,loopTNR}.
Even in tensor renormalization methods with the global optimization, such as HOSRG, short-range correlations are not removed completely.
Such a problem also occurs in our proposed algorithm.
Thus, catching critical phenomena with better accuracy may eventually require the introduction of entanglement filtering techniques \cite{loopTNR, Skeltonization, Branching, GILT, FET}.
In particular, the loop entanglement filtering \cite{loopTNR} and the full environment truncation \cite{FET} seem to fit well with the network structure shown in Fig.~\ref{fig:CTMTRG_cont2}(b).
Moreover, entanglement filtering using the environment may remove short-range correlations more efficiently.
However, this is out of the scope of the present work and remains for future work.

\begin{acknowledgments}
The authors would like to thank K.~Harada, and T.~Okubo for valuable discussions.
The computation in this work was partially executed on computers at the Supercomputer Center, the Institute for Solid State Physics, the University of Tokyo.
This research was supported by MEXT as part of the ``Exploratory Challenge on Post-K Computer'' (Challenge of Basic Science---Exploring Extremes through Multi-Physics and Multi-Scale Simulations) and by JSPS KAKENHI Grants No. JP19H01809 and No. JP20K03780.
\end{acknowledgments}
  
\bibliographystyle{apsrev4-2}
\bibliography{main}

\end{document}